\documentclass[twocolumn,showpacs,preprintnumbers,amsmath,amssymb,superscriptaddress]{revtex4}

\usepackage{graphicx}
\usepackage{dcolumn}
\usepackage{bm}

\begin{document}


\title{Measurement of \mbox{ \boldmath 
$\sigma_{\Lambda^{0}_{b}}/\sigma_{\bar{B}^0} \times {\cal{B}} \left( \Lambda^{0}_{b}\to \Lambda^{+}_{c}\pi^{-} \right) / {\cal{B}} \left( \bar{B}^0 \to D^{+} \pi^{-} \right)$ }
in \mbox{\boldmath $p\bar{p}$} Collisions at 
\mbox{\boldmath $\sqrt{s}=1.96$}~TeV 
}

\date{\today}

\begin{abstract}

We present the first observation of the baryon decay
$\Lambda^{0}_{b}\to\Lambda^{+}_{c}\pi^{-}$ followed by 
$\Lambda_c^+\to p K^-\pi^+$ in $106$~pb$^{-1}$  $p\bar{p}$ 
collisions at $\sqrt{s} = 1.96$~TeV in the CDF experiment.
In order to reduce systematic error, the measured rate 
for $\Lambda^{0}_{b}$ decay is normalized to the
kinematically similar meson decay $\bar{B}^0\to D^{+} \pi^{-}$ followed
by $D^+\to\pi^+K^-\pi^+$.
We report
the ratio of production cross sections ($\sigma$) times the ratio of
branching fractions (${\cal B}$) for the momentum region integrated above 
$p_T > 6$~GeV/$c$ and pseudorapidity range $|\eta| < 1.3$:
$\sigma ( p\bar{p} \rightarrow \Lambda^{0}_{b} X) / \sigma ( p\bar{p} \rightarrow\bar{B}^0 X) 
\times {\cal{B}} (\Lambda^{0}_{b} \to \Lambda^{+}_{c} \pi^{-}) / {\cal{B}} (\bar{B}^0 \to D^{+} \pi^{-}) = 0.82 \pm 0.08\mbox{ (stat) }\pm 0.11\mbox{ (syst) } \pm 0.22 \mbox{ (} {\cal B}(\Lambda^{+}_{c}\to p K^-\pi^+))$.
\end{abstract}



\affiliation{Institute of Physics, Academia Sinica, Taipei, Taiwan 11529, Republic of China} 
\affiliation{Argonne National Laboratory, Argonne, Illinois 60439} 
\affiliation{Institut de Fisica d'Altes Energies, Universitat Autonoma de Barcelona, E-08193, Bellaterra (Barcelona), Spain} 
\affiliation{Baylor University, Waco, Texas  76798} 
\affiliation{Istituto Nazionale di Fisica Nucleare, University of Bologna, I-40127 Bologna, Italy} 
\affiliation{Brandeis University, Waltham, Massachusetts 02254} 
\affiliation{University of California, Davis, Davis, California  95616} 
\affiliation{University of California, Los Angeles, Los Angeles, California  90024} 
\affiliation{University of California, San Diego, La Jolla, California  92093} 
\affiliation{University of California, Santa Barbara, Santa Barbara, California 93106} 
\affiliation{Instituto de Fisica de Cantabria, CSIC-University of Cantabria, 39005 Santander, Spain} 
\affiliation{Carnegie Mellon University, Pittsburgh, PA  15213} 
\affiliation{Enrico Fermi Institute, University of Chicago, Chicago, Illinois 60637} 
\affiliation{Joint Institute for Nuclear Research, RU-141980 Dubna, Russia} 
\affiliation{Duke University, Durham, North Carolina  27708} 
\affiliation{Fermi National Accelerator Laboratory, Batavia, Illinois 60510} 
\affiliation{University of Florida, Gainesville, Florida  32611} 
\affiliation{Laboratori Nazionali di Frascati, Istituto Nazionale di Fisica Nucleare, I-00044 Frascati, Italy} 
\affiliation{University of Geneva, CH-1211 Geneva 4, Switzerland} 
\affiliation{Glasgow University, Glasgow G12 8QQ, United Kingdom} 
\affiliation{Harvard University, Cambridge, Massachusetts 02138} 
\affiliation{Division of High Energy Physics, Department of Physics, University of Helsinki and Helsinki Institute of Physics, FIN-00014, Helsinki, Finland} 
\affiliation{University of Illinois, Urbana, Illinois 61801} 
\affiliation{The Johns Hopkins University, Baltimore, Maryland 21218} 
\affiliation{Institut f\"{u}r Experimentelle Kernphysik, Universit\"{a}t Karlsruhe, 76128 Karlsruhe, Germany} 
\affiliation{High Energy Accelerator Research Organization (KEK), Tsukuba, Ibaraki 305, Japan} 
\affiliation{Center for High Energy Physics: Kyungpook National University, Taegu 702-701; Seoul National University, Seoul 151-742; and SungKyunKwan University, Suwon 440-746; Korea} 
\affiliation{Ernest Orlando Lawrence Berkeley National Laboratory, Berkeley, California 94720} 
\affiliation{University of Liverpool, Liverpool L69 7ZE, United Kingdom} 
\affiliation{University College London, London WC1E 6BT, United Kingdom} 
\affiliation{Massachusetts Institute of Technology, Cambridge, Massachusetts  02139} 
\affiliation{Institute of Particle Physics: McGill University, Montr\'{e}al, Canada H3A~2T8; and University of Toronto, Toronto, Canada M5S~1A7} 
\affiliation{University of Michigan, Ann Arbor, Michigan 48109} 
\affiliation{Michigan State University, East Lansing, Michigan  48824} 
\affiliation{Institution for Theoretical and Experimental Physics, ITEP, Moscow 117259, Russia} 
\affiliation{University of New Mexico, Albuquerque, New Mexico 87131} 
\affiliation{Northwestern University, Evanston, Illinois  60208} 
\affiliation{The Ohio State University, Columbus, Ohio  43210} 
\affiliation{Okayama University, Okayama 700-8530, Japan} 
\affiliation{Osaka City University, Osaka 588, Japan} 
\affiliation{University of Oxford, Oxford OX1 3RH, United Kingdom} 
\affiliation{University of Padova, Istituto Nazionale di Fisica Nucleare, Sezione di Padova-Trento, I-35131 Padova, Italy} 
\affiliation{University of Pennsylvania, Philadelphia, Pennsylvania 19104} 
\affiliation{Istituto Nazionale di Fisica Nucleare Pisa, Universities of Pisa, Siena and Scuola Normale Superiore, I-56127 Pisa, Italy} 
\affiliation{University of Pittsburgh, Pittsburgh, Pennsylvania 15260} 
\affiliation{Purdue University, West Lafayette, Indiana 47907} 
\affiliation{University of Rochester, Rochester, New York 14627} 
\affiliation{The Rockefeller University, New York, New York 10021} 
\affiliation{Istituto Nazionale di Fisica Nucleare, Sezione di Roma 1, University di Roma ``La Sapienza," I-00185 Roma, Italy} 
\affiliation{Rutgers University, Piscataway, New Jersey 08855} 
\affiliation{Texas A\&M University, College Station, Texas 77843} 
\affiliation{Istituto Nazionale di Fisica Nucleare, University of Trieste/\ Udine, Italy} 
\affiliation{University of Tsukuba, Tsukuba, Ibaraki 305, Japan} 
\affiliation{Tufts University, Medford, Massachusetts 02155} 
\affiliation{Waseda University, Tokyo 169, Japan} 
\affiliation{Wayne State University, Detroit, Michigan  48201} 
\affiliation{University of Wisconsin, Madison, Wisconsin 53706} 
\affiliation{Yale University, New Haven, Connecticut 06520} 
\author{A.~Abulencia}
\affiliation{University of Illinois, Urbana, Illinois 61801}
\author{D.~Acosta}
\affiliation{University of Florida, Gainesville, Florida  32611}
\author{J.~Adelman}
\affiliation{Enrico Fermi Institute, University of Chicago, Chicago, Illinois 60637}
\author{T.~Affolder}
\affiliation{University of California, Santa Barbara, Santa Barbara, California 93106}
\author{T.~Akimoto}
\affiliation{University of Tsukuba, Tsukuba, Ibaraki 305, Japan}
\author{M.G.~Albrow}
\affiliation{Fermi National Accelerator Laboratory, Batavia, Illinois 60510}
\author{D.~Ambrose}
\affiliation{Fermi National Accelerator Laboratory, Batavia, Illinois 60510}
\author{S.~Amerio}
\affiliation{University of Padova, Istituto Nazionale di Fisica Nucleare, Sezione di Padova-Trento, I-35131 Padova, Italy}
\author{D.~Amidei}
\affiliation{University of Michigan, Ann Arbor, Michigan 48109}
\author{A.~Anastassov}
\affiliation{Rutgers University, Piscataway, New Jersey 08855}
% ...
\collaboration{CDF Collaboration}
\noaffiliation

\maketitle


Weak decays of baryons
containing $b$ quarks are a good laboratory for testing
the Heavy Quark Effective Theory (HQET)~\cite{Falk:1998wk}.
The $\Lambda^{0}_{b}$ baryon is the ground state of the $udb$ quark 
system, and, in
the heavy quark limit, the light degrees of freedom are in the state of 
zero total angular momentum~\cite{Falk:1992ws}.
%
%
Fully hadronic $b\to~c\bar{u}d$
transitions are more complicated in baryons than in mesons
because there are diagrams which are not present in the decays of 
the latter.  
Various extensions of HQET have been used to evaluate the
$\Lambda^{0}_{b}\to\Lambda^{+}_{c}\pi^{-}$ decay
rate~\cite{Guo:1998at}, but the predictions vary over a large
range.
However, in Soft
Collinear Effective Theory (SCET)~\cite{ref:SCET}, all tree-level
amplitudes can be properly evaluated, 
resulting in
an explicit prediction for
the ratio of branching fractions
${\cal B}(\Lambda^{0}_{b}\to\Lambda_c^+\pi^-)/{\cal B}(\bar{B}^0\to D^{+}\pi^-) \approx 1.7$~\cite{Leibovich:2003tw}.
The decays of $\Lambda^{0}_{b}$ are also interesting 
because they may provide a means to determine
CKM matrix elements with different systematic uncertainties
than the determinations from the decays of $B$ 
mesons~\cite{Dunietz:1992ti}.

This is the first reconstruction of a hadronic
decay of a $b$ baryon at a hadron collider that does not use a $J/\psi$
in the final state.  
In addition, our sample has more than an order of 
magnitude more events than any previous sample of fully reconstructed 
$\Lambda^{0}_{b}$ decays, and, for the same luminosity, is 
about five times larger than a sample of $\Lambda_b \to J/\psi\Lambda$
decays.
Since $b$ baryons are not
produced at the $B$ factories operating at the $\Upsilon(4S)$ resonance,
studying them comprises a unique facet of the $B$ physics program
at Collider Detector at Fermilab (CDF~II)~\cite{Acosta:2004yw}. 
In particular, a large sample of fully reconstructed $\Lambda^{0}_{b}$
decays would allow CDF to study other properties of $b$~baryons, {\it
e.g.}, to measure the lifetime of $\Lambda^{0}_{b}$, and also to search for
decays of heavier $b$ baryons such as 
$\Sigma_b^\pm \to \Lambda^{0}_{b} \pi^\pm$.

This paper presents a measurement of a ratio of $\Lambda^{0}_{b}$ and
$\bar{B}^0$ branching fractions multiplied by the ratio of production
cross-sections,
\begin{equation}
R \equiv
\frac{\sigma ( p\bar{p} \rightarrow \Lambda^{0}_{b} X)}
{\sigma ( p\bar{p} \rightarrow\bar{B}^0 X  )}
\times
\frac
{ {\cal{B}} (\Lambda^0_{b} \to \Lambda^{+}_{c} \pi^{-}) }
{ {\cal{B}} (\bar{B}^0 \to D^{+} \pi^{-})           }
{\rm ,}
	\label{eq:def_R}
\end{equation}
where the $\sigma$ quantities 
are the cross-sections for
$\Lambda^{0}_{b}$ and $\bar{B}^0$ production in the
pseudorapidity range $|\eta|<1.3$
with momentum in the transverse plane, $p_T$, 
above $6$~GeV$/c$~\cite{det_coord}.

The quantity $R$ compares the branching fractions of the topologically 
similar, 
fully reconstructed decays $\Lambda^{0}_{b} \to \Lambda^{+}_{c} \pi^{-}$
and $\bar{B}^0 \to D^{+} \pi^{-}$, where the charmed hadrons
decay via similar three-body channels 
$\Lambda^{+}_{c} \to p K^{-} \pi^{+}$ and 
$D^+ \to \pi^+ K^- \pi^+ $~\cite{conj_modes}.  To extract a branching ratio,
a good understanding of the absolute reconstruction efficiency is
needed.  However, to obtain $R$ only the ratio of the reconstruction
efficiencies of two topologically and kinematically similar decay modes
is evaluated, significantly reducing the systematic errors
on the measured quantity.

The upgraded CDF~II detector is well-suited for the
detailed study of weak decays of heavy baryons.
In particular, the advent of the Silicon Vertex Trigger (SVT)~\cite{Ashmanskas:2003gf},
which uses precise position measurements to select
events containing weakly decaying heavy hadrons, allows CDF~II 
to collect many hadronic decay modes of heavy baryons for the
first time. 
This measurement is performed using a $106$~pb$^{-1}$ sample of
$p\bar{p}$ collisions collected by CDF~II between February 2002 and
June 2003. This data sample corresponds to
$\sim 10^{9}$ $b$ hadron decays produced in the central detector region. 
A full description
of the CDF~II detector can be found elsewhere~\cite{Acosta:2004yw}. The
detector components pertinent to this analysis are the silicon
microstrip vertex detector (SVX~II)~\cite{Sill:2000zz}, the
drift-chamber central tracker (COT)~\cite{Affolder:2003ep} and a
three-tiered trigger system (Levels~1, 2, and 3).  The five double-sided
layers of the SVX~II used in this analysis provide up to $10$ position
measurements. 
Of these, up to five are in the $r$-$\phi$~\cite{det_coord} plane
(each precise to about $15\;\mu$m), three are longitudinal, and two
are small-angle-stereo. The $\phi$ strips are parallel to the
$z$-axis, longitudinal strips are inclined at $90^{\circ}$, and the
small-angle-stereo strips are inclined at $1.2^{\circ}$.
The SVX~II spans the
radii between $2.5$ and $10.6$ cm, and covers the pseudorapidity range
$|\eta| < 2$.
The COT has
96 measurement layers between the radii of $40$ and $137$ cm. These
are organized into alternating axial and small-angle-stereo
($2^{\circ}$) super-layers.  
The COT has a smaller pseudorapidity
coverage ($|\eta| < \; 1.3$) than the SVX~II. 
Both tracking detectors are immersed in a $1.41$~T
magnetic field parallel to the $z$ axis.

This analysis in particular relies on the
SVT, which operates as a part of 
the Level 2 trigger system.
The trigger
makes it possible to select events at a rate of
$\sim 100\; {\rm Hz}$ from the $\sim 1\; {\rm MHz}$ interaction rate. The components of the
three level trigger system pertinent to this measurement are the
Extremely Fast Tracker (XFT) at Level~1 and the SVT at Level~2. 
The XFT uses four axial super-layers of
the COT to find tracks with $p_T > 1.5$~GeV$/c$.
The SVT
combines the XFT measurement with $r-\phi$ hits from the SVX~II
detector.  
The track finding is performed using a large
look-up table of hit patterns.  The found track candidates 
are fitted for 
curvature, angle projected onto the transverse plane
and impact-parameter~\cite{det_track_params}.  
The impact-parameter measurement allows the selection of long-lived
particles in the trigger decision.

The signal ($\Lambda^{0}_{b}$) and normalization ($\bar{B}^0$) events
are collected using the same trigger. At Level~1, two
tracks must satisfy $p_T > 2.0$~GeV$/c$, a
scalar sum of transverse momenta $p_{T1} + p_{T2}> 5.5$~GeV$/c$, and
an angular separation projected onto the transverse plane of
$\Delta\phi < 135^{\circ}$. At Level~2 the transverse momentum cuts
are repeated, and it is required that each track has impact parameter
$d_{0} > 120$~$\mu$m, with an angular separation between the tracks
projected onto the transverse plane of
($2^{\circ}<\Delta\phi<90^{\circ}$). 
Finally, the distance evaluated in the transverse plane from the primary vertex to the
two-track intersection point must be greater than 
$200$~$\mu$m.

Additional criteria are imposed on
the triggered sample in order to
reject as many background events as possible while keeping most of the
signal.
To reconstruct $\Lambda^{0}_{b} \to \Lambda^{+}_{c}\pi^{-}$,
minimum acceptable COT and SVX hit requirements are imposed, and all
valid four track combinations are considered.
The $p_T$ of the proton candidate from the $\Lambda^{+}_{c}$ and the
$\pi^{-}$ candidate from the $\Lambda^{0}_{b}$ must be
greater than $2.0$~GeV$/c$, which strongly favors these
particles to be the two which caused the event to pass the
trigger. The $p_T$ of the proton candidate must be larger
than the $p_T$ of the $\pi^{+}$ candidate from the
$\Lambda^{+}_{c}$. The $p_T$ of the $\Lambda^{0}_{b}$ and
$\Lambda^{+}_{c}$ candidates must be greater than 
$7.5$~GeV$/c$ and $4.5$~GeV$/c$, respectively.
In Equation~(\ref{eq:def_R}), $\sigma_{\Lambda^{0}_{b}}$ and
$\sigma_{\bar{B}^0}$ are defined for $p_T>6$~GeV$/c$.
The events in the data sample must satisfy $p_T>7.5$~GeV$/c$,
and the difference is accounted for by using the Monte
Carlo simulation based on the $p_T$ distributions of both
$\Lambda_b^0$ and $\bar{B}^0$ measured in data~\cite{Yu:2005hw}.

Each of the unstable particles ($\Lambda^{0}_{b}$, $\bar{B}^{0}$,
$\Lambda^{+}_{c}$, and $D^{+}$) is reconstructed by considering all
valid combinations of tracks and requiring them to satisfy the decay
hypothesis.  The charmed hadrons $\Lambda_c^+$ and $D^+$ are
reconstructed first: each triplet of tracks that satisfies the
selection criteria (detailed below) is constrained to pass through the
same point, called the decay vertex.  The decay vertex is determined
by varying the track parameters of the stable daughters within their
uncertainties to minimize the $\chi^2$. 
The $\Lambda_c^{+}(D^{+})$ candidate is then combined with a fourth
track to form a $\Lambda^{0}_{b}(\bar{B}^{0})$ candidate.  
The full topology of the decay is then imposed in another kinematic
fit, resulting in a simultaneous measurement of the 
$\Lambda^{0}_{b}$ ($\bar{B}^0$) and $\Lambda^{+}_{c}$ ($D^+$) vertices.

Using these measurements, the reconstructed invariant
mass of the $\Lambda^{+}_{c}$ must be between $2.269$~GeV$/c^{2}$
and $2.301$~GeV$/c^{2}$.
Other selection criteria rely on $L_{xy}$, the projection onto
the $x$-$y$ (transverse) plane of the decay length measured
from the production vertex to the decay point;
the production vertex is estimated by the
position of the beamline averaged over each run calculated for
the $z$ coordinate of the secondary vertex.
A product of the proper decay time
and the speed of light, $ct$, is also used.  It is derived from $L_{xy}$: 
$ct = (\vec{L}_{xy} \cdot \hat{p}_T)(mc/p_T)$, where
$\vec{L}_{xy}$ is the decay vector of $\Lambda^{0}_{b}$ ($\bar{B}^0$)
projected onto the $x$-$y$ plane, $p_T$ is the transverse momentum,
 $\hat{p}_T$ is the unit vector in the
direction of the transverse momentum,
and $m$ is
the world average mass of the $\Lambda^{0}_{b}$ ($\bar{B}^0$).  
In order to suppress the combinatorial background from the interaction
point, we
impose $ct(\Lambda^{0}_{b}) > 225$~$\mu$m (compared
to the $b$~baryon mean decay length of $368 \pm 24$~$\mu$m~\cite{PDG}).
Calculated relative to the $\Lambda^{0}_{b}$ decay point, $ct(\Lambda^+_c)$
must be $ct(\Lambda^+_c) > -65$~$\mu$m.
For a true $\Lambda_c^+$, a small negative $ct$ may arise due 
to resolution effects; on the other hand, the $ct(\Lambda_c^+)$ of
combinatorial background candidates may have large negative values.
The distance of closest approach in the transverse plane of the
trajectory of the $\Lambda^{0}_{b}$ candidate to the primary vertex must
be less than $85 \; \mu m$. The $\chi^{2}_{xy}$ of the
$\Lambda^{0}_{b}$ and $\Lambda^{+}_{c}$ kinematic fits are required to
be less than $30$ and $20$ respectively, where $\chi^{2}_{xy}$ is a
$\chi^{2}$-like goodness-of-fit quantity using only
the track parameters in the transverse plane.

\begin{figure}
\begin{center}
\includegraphics[height=7cm]{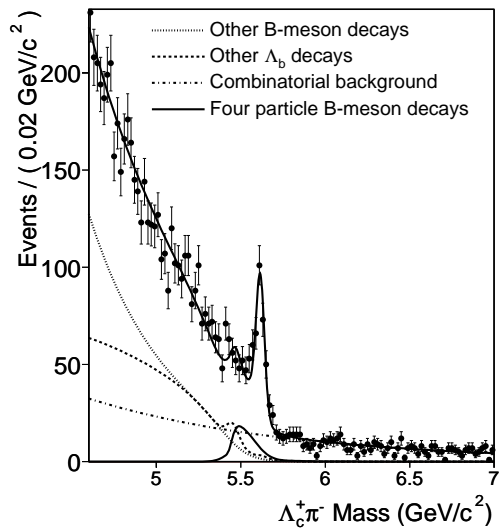}
\caption{$\Lambda^{0}_{b} \to \Lambda^{+}_{c} \pi^{-}$ yield with binned 
	likelihood mass fit. The background shapes are defined in the text.}
\label{lb_fit}
\end{center}
\end{figure}

\begin{figure}
\begin{center}
\includegraphics[height=7cm]{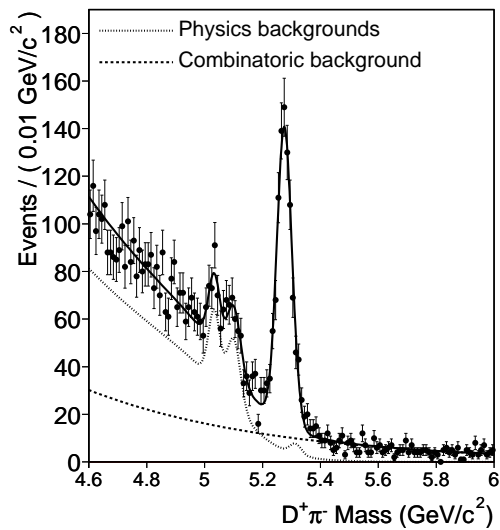}
\caption{$\bar{B}^0 \to D^{+} \pi^{-}$ yield with binned likelihood 
	fit. The background shapes are defined in the text.}
\label{b0_fit}
\end{center}
\end{figure}

The normalization mode ($\bar{B}^0 \to D^{+} \pi^{-}$)
is reconstructed using selection criteria identical to those of the signal
mode, except for a different invariant mass requirement 
for the $D^{+}$ candidate and no
analogy to the $p_T(p) > p_T(\pi^+)$ cut.  
The $D^{+}$ candidate invariant mass must be between 
$1.848$~GeV$/c^{2}$ and $1.888$~GeV$/c^{2}$.  The distributions of the
invariant mass of $\Lambda^{0}_{b}\to\Lambda^{+}_{c} \pi^{-}$ and
$\bar{B}^0 \to D^{+} \pi^{-}$ candidates
are shown in Figures~\ref{lb_fit} and~\ref{b0_fit}, respectively.

Figure~\ref{lb_fit} shows the binned likelihood fit to the invariant
mass distribution of $\Lambda^{0}_{b}$ candidates.  The large Gaussian 
peak at $5.6$~GeV/$c^2$ is the signal. The dash-dotted curve
corresponds to the exponential combinatorial background. This component
is constrained by the data in the invariant mass region above the
$\Lambda^{0}_{b}$ mass.  
The small asymmetric peak at
$5.5$~GeV/$c^2$ (solid line) corresponds to contributions from
fully reconstructed $B$-meson decays resulting in a final state
with four tracks, where at least one track is misidentified. 
This shape is obtained using
a full detector simulation of these modes. It is consistent
with the shape of $\bar{B}^0\to D^+\pi^-$ candidates found in the
$\Lambda^{0}_{b}$ sample.
The dotted and dashed curves correspond to all the other $B$-meson and
$\Lambda^{0}_{b}$ backgrounds, respectively.  These shapes are determined
from a large parametric Monte Carlo sample which includes all 
known decays of $B^+$,
$B^0$, and $B_s$ and $\Lambda_b^0$ hadrons.
Finally, there is a very small Gaussian distribution (not shown) from the
Cabibbo-suppressed mode $\Lambda^{0}_{b}
\to\Lambda^{+}_{c} K^{-}$ (fixed to an expected $8\%$ of the signal 
yield~\cite{Abe:2001wa}). The total distribution is 
$F_{{\rm tot}} = 
G_{{\rm signal}} + 
E_{{\rm comb}} +
F_{{\rm four-track}} + 
F_{{\rm other-B}} + 
F_{{\rm other-}{\Lambda^{0}_{b}}} +
G_{\Lambda^{+}_{c} K^{-}},$
where $G$ indicates a Gaussian distribution, $E$ indicates an exponential, and $F$
indicates a more complicated functional form.

In the fit, the width of $\Lambda^{0}_{b}$ signal is fixed to
$26.4$~MeV/$c^2$, obtained by scaling the $\bar{B}^0$ width in data
by the ratio of widths from the Monte Carlo simulation.
%
%
The relative contribution of each background type
in the fit is guided by two constraints: the first describes the
normalization of $F_{\rm four-track}$ relative to $F_{\rm other-B}$
({\it i.e.}, $N_{\textrm{four-track}}/N_{\textrm{other-B}}$), where $N$ is
number of events; the second describes the normalization of
$F_{{\rm other-}{\Lambda^{0}_{b}}}$ relative to $\left( F_{\rm other-B} + F_{\rm four-track}\right)$, 
({\it i.e.}, 
$N_{\textrm{other }\Lambda^{0}_{b}}/ (N_{\textrm{four-track}} + N_{\textrm{other-B}})$).
The value of each constraint is inferred from the relative
abundance of the background types in the large parametric Monte Carlo sample. 
The value of the $\left(N_{\textrm{four-track}}/N_{\textrm{other-B}}\right)$
constraint is checked by reconstructing the $\bar{B}^0 \to D^+\pi^-$ mode 
among $\Lambda_c^+\pi^-$ candidates from the region of the invariant
mass corresponding to the $\Lambda_b^0$ signal.  
The total $\chi^{2}$ of this fit is $80.6$ for $88$ degrees of freedom, 
corresponding to the fit probability
of $70\%$.

Figure~\ref{b0_fit} shows the mass fit for the $D^+\pi^-$ candidates.
The large Gaussian peak at $5.27$~GeV$/c^2$ is the $\bar{B}^0$
signal. The dashed curve corresponds to the exponential combinatorial
background.  The dotted curve corresponds to the background from
decays of other $b$ hadrons. 
The total $\chi^{2}$ of this fit is $70.9$ for $94$ degrees of freedom, 
corresponding to the fit probability
of $96\%$.

The quantity $R$ defined in equation~(\ref{eq:def_R}) is calculated
from the signal yield according to
\begin{equation}
R=
\frac
{
N_{\Lambda^{0}_{b}}
}
{
N_{\bar{B}^0}
}
\times
\frac
{
{\cal{B}}(D^+ \to \pi^+ K^- \pi^+)
}
{
{\cal{B}}(\Lambda^{+}_{c} \to p K^{-} \pi^{+})
}
\times
\frac
{
\epsilon ( \bar{B}^0 )
}
{
\epsilon ( \Lambda^{0}_{b} )
},\label{calc_qty}
\end{equation}
where the first factor is the ratio of observed signal yields, the
second factor is the ratio of the daughter branching fractions~\cite{PDG}, 
and the third
factor is the ratio of reconstruction efficiencies calculated from the
Monte Carlo. The signal yields are $N_{\Lambda^{0}_{b}} = 214\pm 19$
and $N_{\bar{B}^0} = 790\pm 32$ respectively. 
Each reconstruction efficiency is defined 
for the $\Lambda^{0}_{b}$($\bar{B}^{0}$) with 
$p_T > 6$~GeV$/c$ and $|\eta| < 1.3$. 
The exact configuration of the CDF~II detector varied over
the course of collecting the data used in the analysis.
However, the ratio of reconstruction efficiencies is stable within
statistical errors across the different periods of running, with the
average value 
$\epsilon(\bar{B}^0)/\epsilon(\Lambda^{0}_{b})=1.65~\pm~0.03$.  
From Equation~(\ref{calc_qty}) we thus obtain $R = 0.82 \pm 0.08\mbox{(stat)}$.

The systematic uncertainty on the measurement of $R$ is dominated by
the error on ${\cal B}(\Lambda^{+}_{c}\to p K^{-} \pi^{+})$, yielding
a relative error of $27\%$~\cite{PDG}.  Since this uncertainty is
independent of our measurement, it is quoted separately.
%
%
Other sources of systematic uncertainty are the imperfect
knowledge of the $\Lambda^{0}_{b}$ lifetime ($^{+5}_{-4}\%$), the
production $p_T$ spectra of both $\Lambda^{0}_{b}$ ($7.6\%$)
and $\bar{B}^0$ ($4\%$), the $\Lambda^{0}_{b}$ polarization ($7\%$), and
the $\Lambda_c^+$ resonant substructure ($1\%$).  The uncertainty
due to the finite size of the Monte Carlo samples is $1.9\%$.
The uncertainty due to the difference between the proton and $\pi^{+}$ 
trigger efficiency is $0.6\%$.
The systematic uncertainties due to the background shapes in the
invariant mass fits are $4.3\%$ for the $\Lambda^{0}_{b}$ and $0.9\%$
for the $\bar{B}^0$.
These uncertainties arise from the lack of detailed knowledge of a
variety of branching fractions contributing to background shapes
that are obtained from the Monte Carlo simulation and fixed in the fit.
To evaluate the uncertainty due to these shapes, the branching
fractions of the largest decay modes contributing to each of the
shapes were varied simultaneously in the simulation, and the shapes were reevaluated.
Uncertainties on the mass resolutions of both $\Lambda^{0}_{b}$ and
$\bar{B}^0$, which are also fixed in the mass fits, contribute
$2.8\%$ and $1.8\%$, respectively.  Finally, the contribution of the
$\Lambda^{0}_{b}\to \Lambda^{+}_{c} K^{-}$ shape is varied by a factor
of $2$, contributing $1.6\%$ to the systematic error.
The total systematic error excluding the uncertainty on
${\cal{B}}(\Lambda_c^+\to p K^-\pi^+)$ is $13.5\%$.

A direct comparison with a theoretical prediction of 
$R_{BR} \equiv {\cal B}(\Lambda_b^0\to\Lambda_c^+\pi^-)/{\cal B}(\bar{B}^0\to D^{(*)+}\pi^-) \approx 1.7$~\cite{Leibovich:2003tw} 
can be performed if one assumes that $\sigma(\Lambda_b)$ and $\sigma(B^0)$
have the same dependence on $p_T$, and then use $f_{\rm baryon}/f_{d}$ 
from high-$p_T$ measurements.  From Ref.~\cite{PDG} we obtain
$f_{\rm baryon}/f_{d} = 0.25 \pm 0.04$, which yields 
$R_{BR} \approx 3.3 \pm 1.2$, in agreement with~\cite{Leibovich:2003tw}.

In summary, we have observed the decay
$\Lambda^{0}_{b}\to\Lambda^{+}_{c}\pi^{-}$ for the first time, and
measured $R = 0.82 \pm 0.08\mbox{(stat)}\pm 0.11\mbox{ (syst)} \pm 0.22 \mbox{ } \mbox{(}{\cal{B}}_{\Lambda^{+}_{c}} \mbox{)}$.  The
overall error is dominated by the large uncertainty on ${\cal B}(\Lambda^{+}_{c} \to p K^-\pi^+)$.  
The $\Lambda^{0}_{b}\to\Lambda^{+}_{c}\pi^{-}$ sample is the largest
$b$-baryon sample in existence, and, once augmented by new data,
can be used for a variety of other $\Lambda_b$ measurements,
including its lifetime and production properties.


We thank the Fermilab staff and the technical staffs of the
participating institutions for their vital contributions. This work
was supported by the U.S. Department of Energy and National Science
Foundation; the Italian Istituto Nazionale di Fisica Nucleare; the
Ministry of Education, Culture, Sports, Science and Technology of
Japan; the Natural Sciences and Engineering Research Council of
Canada; the National Science Council of the Republic of China; the
Swiss National Science Foundation; the A.P. Sloan Foundation; the
Bundesministerium f\"ur Bildung und Forschung, Germany; the Korean
Science and Engineering Foundation and the Korean Research Foundation;
the Particle Physics and Astronomy Research Council and the Royal
Society, UK; the Russian Foundation for Basic Research; the Comisi\'on
Interministerial de Ciencia y Tecnolog\'{\i}a, Spain; in part by the
European Community's Human Potential Programme under contract
HPRN-CT-2002-00292; and the Academy of Finland.


\begin{thebibliography}{99}
\bibitem{det_coord} CDF~II uses a cylindrical coordinate system 
with the $z$-axis along the nominal beam-line. The transverse plane
$(x,y)$ is perpendicular to the $z$-axis.  Azimuthal angle, $\phi$,
is measured from the $x$-axis.  Polar angle, $\theta$, is measured 
from the $z$-axis.  Pseudorapidity is defined as
$\eta \equiv \tanh^{-1}(\cos \theta)$.
Transverse momentum, $p_T$, is the component 
of the particle's momentum projected onto the transverse plane.
\bibitem{conj_modes} Inclusion of the respective charge
conjugate modes is assumed 
throughout this paper.  Specifically, 
$\bar{\Lambda}^0_{b} \to \Lambda^{-}_{c} \pi^{+}$, 
$B^0 \to D^{-} \pi^{+}$, $\Lambda^{-}_{c} \to \bar{p} K^{+} \pi^{-}$,
and $D^{-} \to \pi^{-} K^{+} \pi^{-}$.
\bibitem{det_track_params}
The curvature of a particle
track is inversely proportional to the transverse momentum
$(p_T)$. The impact-parameter of a particle track is defined as the
distance of closest approach of the particle track to the primary
vertex in the transverse plane.
\end{thebibliography}
\end{document}